\documentclass[12pt]{article}
\usepackage{times}
\usepackage{geometry}
\usepackage[utf8]{inputenc}
\usepackage{enumitem,amssymb}
\usepackage{ragged2e}
\usepackage[dvipsnames]{xcolor}
\newlist{thematic}{itemize}{8}
\setlist[thematic]{label=$\square$}
\usepackage{times}
\usepackage{geometry}
\usepackage[utf8]{inputenc}
\usepackage{enumitem,amssymb}
\usepackage{ragged2e}
\usepackage[dvipsnames]{xcolor}
\setlist[thematic]{label=$\square$}
\usepackage{pifont}

\newcommand{\arcsec}{\mbox{$^{\prime \prime}$}}
\usepackage{natbib}
\usepackage{graphicx}
\newcommand{\emilcheckbox}{\ensuremath{\leavevmode\rlap{$\checkmark$}\square}}
\newcommand{\emilnoncheckbox}{\ensuremath{\square}}

\newcommand{\ii}{{\sc ii}}
\newcommand{\iii}{{\sc iii}}
\newcommand{\W}{$\lambda$}

\begin{document}
{\raggedright
\huge
Astro2020 Science White Paper \linebreak

Radial Metallicity Gradients in Star-Forming Galaxies \linebreak
\normalsize

{\bf Thematic Areas:} \\
\emilnoncheckbox\quad   Planetary Systems \emilnoncheckbox Star and Planet Formation\\
\emilnoncheckbox\quad   Formation and Evolution of Compact Objects  \emilcheckbox Resolved Stellar Populations and their Environments\\
\emilnoncheckbox\quad   Cosmology and Fundamental Physics  \emilcheckbox Stars and Stellar Evolution \\
\emilcheckbox\quad   Galaxy Evolution   \emilnoncheckbox Multi-Messenger Astronomy and Astrophysics \\

\bigskip
\textbf{Principal Author:}

Letizia Stanghellini
 \linebreak						
National Optical Astronomy Observatory
 \linebreak
lstanghellini@noao.edu
 \linebreak
(520) 318 8205
 \linebreak
 
\textbf{Co-authors:} 
\linebreak
 Danielle Berg, Ohio State University \linebreak
 Fabio Bresolin, Institute for Astronomy, University of Hawaii \linebreak 
 Katia Cunha, University of Arizona\linebreak
 Laura Magrini, INAF Osservatorio Astrofisico di Arcetri \linebreak

\begin{abstract}
Spiral star-forming galaxies are complex astrophysical objects whose baryonic component is dominated by the disk, where most of the star formation resides. The metallicity in the disk is not uniform, and it usually decreases with the distance to the galaxy center, in the so-called radial metallicity gradient. Radial metallicity gradients have been successfully used to set important constraints on galaxy formation and their chemical evolution. This paper focuses on the implications of radial metallicity gradients measured with different probes for a variety of galaxies, and on the foreseen advances in this field in the astronomical landscape of the 2020s.
\end{abstract}

}

\clearpage
\section{Introduction}
Radial metallicity gradients have been successfully used to set important constraints on the chemical evolution of galaxies.  In spiral, star-forming galaxies, it is generally possible to determine a radial metallicity gradient, with metallicity decreasing from inside out. This has been known for a long time, starting with the pioneering work by Aller (1942), and later by Searle (1971), and Pagel \& Edmunds (1981). The metallicity gradient carries the signature of star formation and evolution, galactic stellar motions, and gas inflows and outflows. 

In recent times, new Galactic and extragalactic survey data gave impulse to the field, and gradients have been used extensively to constrain and interpret star-forming galaxy evolution. The study of metallicity gradients in Galactic populations, both from field stars and open clusters, has entered an exciting stage with vast amounts of spectroscopic data being obtained by the large ongoing Galactic surveys, such as Gaia-ESO, APOGEE, GALAH, and others. These ever increasing large data sets are providing the needed statistics to enable us to answer important questions that impact any understanding of Galaxy formation and evolution, for instance, how metallicity gradients compare and connect with the bulge and halo populations, the importance of azimuthal variations, and whether the chemical tagging paradigm (i.e., each open cluster has a chemical abundance imprint that is unique) holds.

Owing to a plethora of dedicated surveys, important galactic archaeology results have impacted the past decade, (see, e.g., Bergemann et al. 2014; Hayden et al. 2014, Magrini et al. 2016). In particular, great investments have been made to study resolved stellar populations, H\,II regions, and planetary nebulae (PNe) in nearby galaxies with the multi-object facilities (see, e.g., Bresolin 2007; Magrini et al. 2009, 2010, Stanghellini et al. 2010; Berg et al. 2015), to surveys of spatially resolved intermediate distance galaxies (see, e.g, S\'anchez et al. 2014), and, finally, to the study of high-redshift galaxies with the infrared facilities on the 8m-class telescopes, often coupled with adaptive optics (see, e.g., Cresci et al. 2010; Jones et al. 2010, 2013, 2015).

However, there are still many open questions concerning metallicity gradients in disk galaxies: their shapes, including possible (multi-) bi-modality; their time evolution; the ideal element to depict the gradient signature, which may be different in the different environment and for different populations; their azimuthal dependence; their dust-to-gas ratio dependence; and biases introduced by diffuse gas and aperture or integration effects.

The field of the radial metallicity gradients, with their implications for our understanding of the formation and evolution of galaxies, is thus a focal point for a diversity of astronomical communities, observational techniques (involving a variety of stellar populations), and different epochs in the evolution of galaxies. 
The current wealth of observational resources, which has been and will continue to be mined to construct and study radial metallicity gradients and their evolution, is supported by an equally impressive theoretical and modeling effort. A new class of chemo-dynamical models, often built in a fully cosmological framework (see, e.g. Gibson et al. 2013, also in Fig. 1; Grand et al. 2015), considers the shape of the radial metallicity gradient and its time evolution as a major observational constraint. The field is of great important to studies of galaxies that span from the Milky Way to lensed galaxies at $z\sim2-3$.

In this paper, we discuss a selection of radial metallicity gradients studied in star-forming galaxies in order to showcase their importance and interdisciplinary nature in galaxy evolution studies. We attempt to give a current assessment of the field, and to review the astrophysical limits set by available instrumentation. We also discuss radial metallicity gradients in view of the future instrumental landscape, in particular, the foreseen advances in this field with the advent of extremely-large (30m-class) telescopes (ELTs).

\section{Radial metallicity gradients in the Milky Way}
\subsection{Emission-line probes}
Planetary nebulae and H\,II regions are both excellent probes of radial metallicity gradients in the Galaxy.  For these emission-line probes, oxygen is the metallicity proxy. At zeroth order, oxygen does not change through low- and intermediate-mass stellar evolution, thus PNe probe the oxygen abundance of their progenitors. 
While H\,II  regions track the latest stellar generations, PNe can explore lookback times from $\sim$1 to 5 Gyr, depending on the initial progenitor mass. Galactic PN and H\,II region oxygen abundances available to date, measured via auroral-line plasma diagnostic (i.e., the direct method) disclose that the radial oxygen gradient steepens with time (Stanghellini \& Haywood 2010, see also Fig. 1). While the process of dating AGB progenitors is fairly successful (e.g., Stanghellini \& Haywood 2018), these gradients suffer the uncertainties of the Galactic PN distance scale. Current Gaia or other parallaxes are available only for small galactocentric distances (R$_{\rm G}<10$ kpc), strongly limiting the gradient analysis. Progress is expected from future Gaia releases, and by moving this type of analysis to nearby (see $\S$3) and distant galaxies.

\subsection{Large spectroscopic surveys of field and clusters} %written by LM
Galactic archaeology is going through a golden age thanks to the connection between
the Gaia mission (Gaia Collaboration et al. 2018) and the many on-going 
ground-based spectroscopic surveys at low/intermediate spectral resolution, such as RAVE  (Steinmetz et al. 2006), SEGUE (Yanny et al. 2009) and LAMOST (Liu et al. 2017) and at high resolution, such as APOGEE (Majewski et al. 2017), Gaia-ESO (Gilmore et al. 2012) and GALAH (Martell et al. 2017).

The presently available databases  already represent a revolution in our knowledge and understanding of the spatial distribution of chemical abundances, allowing us to disentangle the 3D-chemical structure at unprecedented scale and detail. 
Thanks to age-dating methods (e.g., asteroseimology, isochrone fitting of star cluster sequences, chemical indicator of age)  we can also discern the time evolution of the chemical structure of the Galaxy, including the shape of the radial metallicity gradient. Comparing observations with theoretical models, we can identify  the role of various processes (such as infall, outflow, radial migration, etc.) acting to shape the geometry, dynamics, and chemistry of the Galaxy. \looseness=-2

\section{Radial metallicity gradients in nearby galaxies} % written by FB

Since the 1970's the determination of radial abundance gradients in nearby galaxies has focused on oxygen, the most abundant of the metals, as well as the most straightforward to measure in the emission-line spectra of extragalactic H\,II regions. 
The vast majority of investigations aiming at measuring abundance gradients of ever larger samples of galaxies, both nearby and at high redshift -- often limited to observations of the strongest spectral lines --  utilize a statistical technique for measuring the gas chemical composition, which is affected by large systematic uncertainties.

 In place of a direct measure of the ionized gas electron temperature, which requires the detection of faint auroral lines -- a difficult task even using 8m-class telescopes for galaxies situated beyond a few Mpc from the Milky Way (Berg et al.~2015) -- `strong-line' methods have been developed (e.g.~Pettini \& Pagel~2004), and are widely adopted in order to derive the abundance gradients of star-forming galaxies. Importantly, all the current generation of IFU-based surveys of thousands of galaxies (CALIFA - S\'anchez et al.~2012; SAMI - Bryant et al.~2015; MaNGA - Bundy et al.~2015) adopt, without exception, one or more of these strong-line methods, often selecting different calibrations. 
This approach is  presently providing a wealth of emission-line data that is shaping our knowledge of galactic abundance gradients. Recent results include the finding of a common O/H gradient scale length among nearby galaxies (S\'anchez et al.~2014), the dependence of the gradient slope on galaxy mass (Belfiore et al.~2017), and the frequent presence of `inverted' gradients in the central regions of spiral disks (S\'anchez-Menguiano et al.~2018). However, the analysis based on strong-line diagnostics introduces systematics in the deduced chemical abundance properties  that are both difficult to quantify and detrimental for our quest to pin down the chemical evolution of the Universe.

\begin{figure}[h!]
\centering
\includegraphics[scale=0.4]{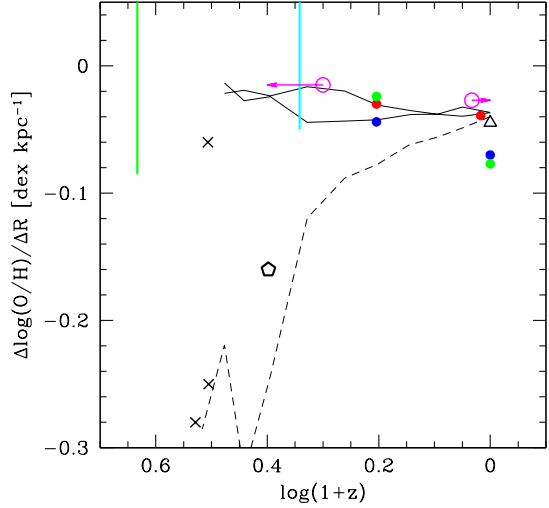}
\caption{Radial oxygen gradient from star-forming galaxies vs. redshift. Black tracks are representative of inside-out chemical galaxy evolution models in a cosmological context: solid with enhanced feedback; dashed with no feedback (Gibson et al. 2013). Symbols represent gradient slopes from different data sets. Open circles: Galactic PNe with old and young progenitors (Stanghellini \& Haywood 2018). Triangle: Galactic H\,II regions (Balser et al. 2011). Red circles: M33 PNe and H\,II region (Magrini et al. 2009, 2010). Blue circles: PNe and H\,II regions in M81 (Stanghellini et al. 2010, 2014). Green circles: NGC300 PNe (Stasinska et al. 2013) and H\,II regions (Bresolin et al. 2009). Pentagon: Yuan et al. (2011). Crosses: Jones et al. (2013). Vertical lines: ranges of gradient slopes from Cresci et al. (2010, green) and S\'anchez et al. (2014, cyan). 
}
\end{figure}
Tests on the reliability of the extragalactic nebular abundances obtained from the widespread use of statistical abundance diagnostics have been carried out in recent years. These involve the use of stellar abundances for individual blue (Bresolin et al.~2016) and red (Davies et al.~2015) supergiant stars, as well as young stellar superclusters (Gazak et al.~2014). A significant advantage of using stellar spectra is the fact that the systematic uncertainties in the metallicities are on the order of 0.15 dex, much smaller than the 0.7 dex currently estimated for strong-line abundances (Kewley \& Ellison~2008). The future use of wide-field IFUs on large telescopes will enable the extension of this complementary approach to derive abundance gradients of galaxies from young stars out to the Virgo Cluster. Such instrumentation will also be of foremost importance to measure the chemical composition of faint H\,II regions that populate the very outskirts of spiral disks, where remarkably shallow or flat gradients are observed (Bresolin et al.~2012). Probing the distribution of metals at galactocentric distances of tens of kpc is critical to examine the evolutionary processes that take place in star-forming galaxies and in the circumgalactic medium.

\section{Planning for the 2020s and beyond: \\
Advances with future instrumentation} 

We expect extraordinary advances in the next decade in the field of radial metallicity gradients. 
Concerning the Galaxy, the enhancement of the multi-object spectrometers (4MOST, de Jong et al. 2012; WEAVE, Dalton et al. 2012; MOONS, Cirasuolo et al. 2011; MSE, Zhang et al. 2016) and, in particular, the next generation of ELTs, will enlarge our view of the Galaxy complementing the photometric and astrometric data from Gaia and LSST with accurate chemical abundances even for the faintest stars, from  medium-resolution MOS observations (in the optical or infrared) or from high resolution spectrographs. 

The advent of ELTs will enable further improvements on abundance systematics, by obtaining nebular abundance gradients for nearby spirals from metal recombination lines, rather than using the traditional collisionally excited lines. 
Historically, abundances based on optical recombination lines are systematically discrepant to those derived from collisionally-excited lines.
This ``abundance discrepancy problem" has existed since the seminal paper of Peimbert \& Costero (1969) 
and our inability to determine an absolute abundance scale persists as one of the most significant challenges 
in the field. To date, the detection of these spectral features has been sporadic in the H\,II regions of nearby galaxies, due to their weakness (Esteban et al.~2009), but the results stress our inadequate understanding of systematic effects in nebular abundances.

\begin{figure}[h!]
\centering
\includegraphics[scale=0.5]{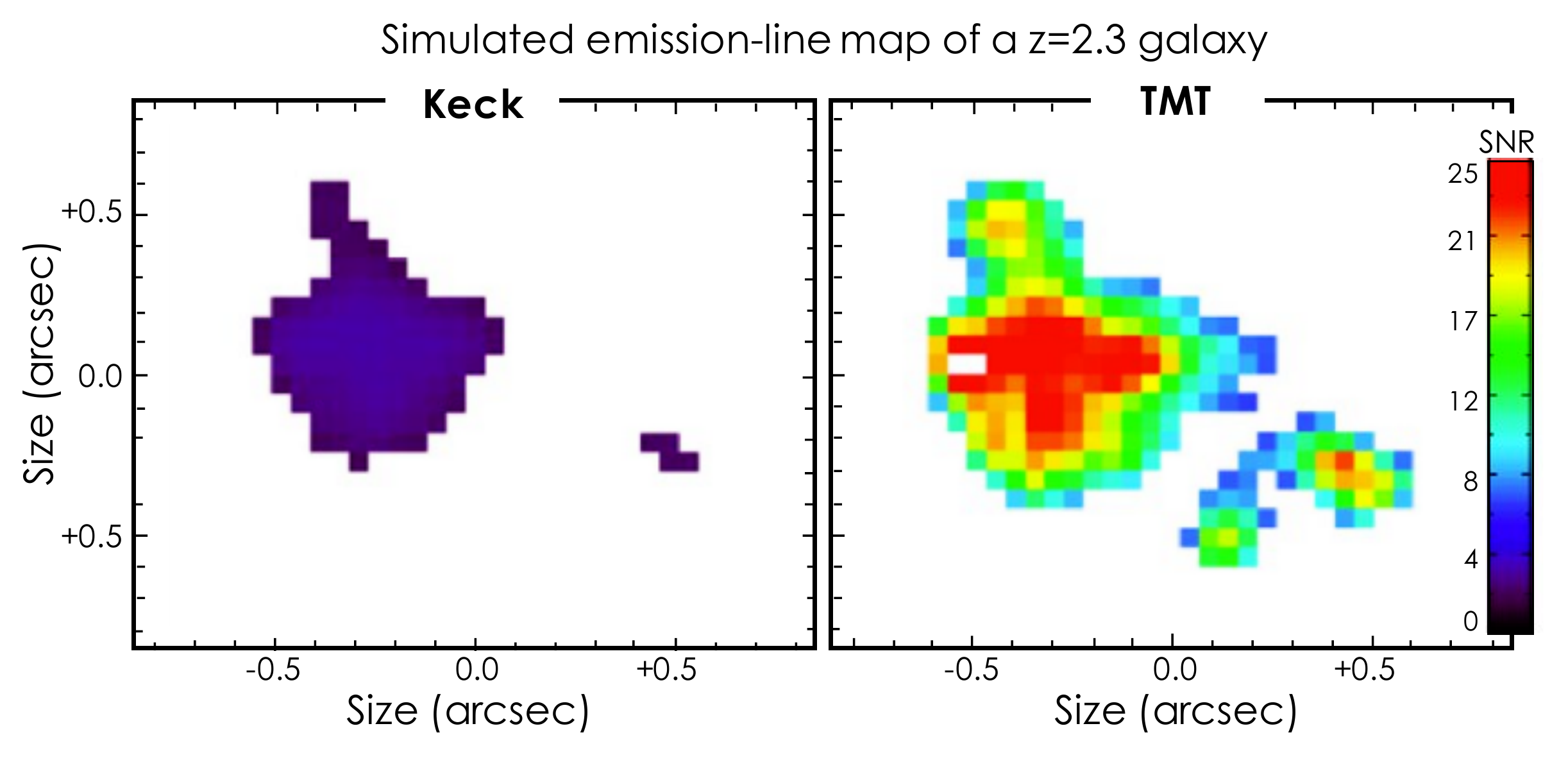}
\caption{
A simulated $z = 2.3$ galaxy, with H$\alpha$ redshifted in to the K-band and spatially 
sampled at 0.05$\arcsec$ per spaxel.
{\it Left:} the nebular emission is only marginally detected (SNR$\sim3$) in the brightest part of the galaxy after a 2 hour integration using a 10m telescope, such as Keck.
{\it Right:} significantly higher S/N ratios will be achieved with the same exposure time with ELTs, such as TMT, revealing a complex structure, with multiple knots, and large variations in emission-line maps across the galaxy. The sensitivity and resolution of ELTs, as shown here, are essential for detecting the suite of rest-frame optical emission lines (i.e., [N~\ii], [O~\iii], [S~\ii]) necessary to measure ionization parameters and chemical abundance patterns across galaxies beyond the local volume
(adapted from Wright et al 2014). }
\end{figure}
Observing abundance gradients in a given galaxy from a variety of objects with different ages, 
such as supergiant stars, H\,II  regions, and Cepheids to trace the present abundance gradient, open clusters
to probe the gas a few Gyr ago when they formed, and PNe to probe a large range of lookback times, allows us to determine the gradient evolution.
Presently, we are limited by our 8m-class telescopes to observing our toolset of abundance tracers in only a handful of the closest galaxies, where even direct abundances in H\,II regions and individual stellar abundances throughout galaxies can only probe $\lesssim10$ Mpc.
The next generation of ELTs will greatly benefit the determination of metallicity gradients,
allowing increased sensitivity and angular resolution delivered by adaptive optics to trace the abundances 
across galaxies out to intermediate redshifts ($z\sim1$).
For instance, a MUSE-like IFU with adaptive optics, which boasts spatial resolutions on the order of 0.05\arcsec,
combined with the photon collecting area of an ELT could easily probe metallicity gradients
on spatial scales of 24 pc in galaxies 100 Mpc away.
Note that the full suite of rest-optical emission lines needed to determine direct nebular oxygen abundances
(i.e., [O~\ii] \W3727, H$\gamma$, [O~\iii] \W4363, H$\beta$, and [O~\iii] \W5007) is within ground-based reach for redshifts of $z\lesssim1$. 
The increased sensitivity will also allow us to accurately trace abundance profiles into the faint outskirts of disk galaxies, providing important implications for disk formation models.

Multi-object spectroscopic instrument associated with ELTs, such as the WFOS for TMT, can make a breakthrough in the field of emission-line probes abundances in star-forming galaxies. With a good sampling of probes within the FoV across spiral disks, spectra of auroral lines could be acquired in minutes, as opposed to the two or more nights needed with current apertures, for PNe at the distance of M81. 
The great advance afforded by the ELT MOS modes is to open up the realm of metallicity gradients in Virgo galaxies,  where the variety of galaxy types can be examined in details for their chemical content though emission-line probes. Direct oxygen abundances in Virgo galaxies can be observed rapidly for any resolved spiral galaxy with an ELT MOS. \looseness=-2

Infrared IFU modes are useful to determine the radial metallicity gradient of redshifted galaxies. For example, direct abundances of H\,II regions in galaxies with $z\sim1.6$ should be possible, since the usual diagnostic lines (i.e., [O~\ii] \W3727, [O~\iii] \W4363, H$\beta$, H$\alpha$, [N~\ii] \W\W6548,6584, and [O~\iii] \W\W4959,5007) are seen in YJH bands at these redshifts. 
Furthermore, IR IFU modes on ELTs will also open up gradient studies in galaxies with $2 < z < 2.6$, where abundance diagnostic lines can be observed in the JHK bands. The ELTs will afford adequate spatial resolution to detect metallicity gradients (see Fig. 2).

\clearpage

\bibliographystyle{plain}
\bibliography{references}

Aller, L. H. 1942, ApJ 95, 52 
\noindent

Balser, D. S., et al. 2011, ApJ 738, 27
\par\noindent

Belfiore, F., et al. 2017, MNRAS, 469, 151
\noindent

%Berg, D., et al. 2012, ApJ 806, 16
%\noindent

%Berg, D., et al. 2013, ApJ 775, 128
%\noindent

Berg, D., et al. 2015, ApJ 806, 16
\noindent

Bergemann, M., et al. 2014, A\&A 565, 89
\noindent

Bresolin, F. 2007, ApJ 656, 186
\noindent

Bresolin, F., et al. 2009, ApJ 700, 309
\noindent

Bresolin, F., et al. 2012, ApJ, 750, 122
\noindent

Bresolin, F., et al. 2016, ApJ, 830, 64
\noindent

Bryant, J. J., et al. 2015, MNRAS, 447, 2857
\noindent

Bundy, K., et al. 2015, ApJ, 798,7
\noindent

Cirasuolo, M., Afonso, J., Bender, R., et al. 2011, The Messenger, 145, 11 
\noindent

Cresci, G., et al. 2010, A\&A 520, 82
\noindent

Dalton, G., Trager, S.~C., Abrams, D.~C., et al. 2012, Proc. SPIE, 8446, 84460P 
\noindent

Davies, B., et al. 2015, ApJ, 806, 21
\noindent

de Jong, R.~S., Bellido-Tirado, O., Chiappini, C., et al. 2012, Proc. SPIE, 8446, 84460T 
\noindent

Esteban, C., et al. 2009, ApJ, 700, 654
\noindent

Gaia Collaboration, Brown, A.~G.~A., Vallenari, A., et al.\ 2018, A\&A, 616, A1 
\noindent

Gazak, J. Z. 2014, ApJ, 787, 142
\noindent

Gilmore, G., Randich, S., Asplund, M., et al. 2012, The Messenger, 147, 25 
\noindent

Gibson, B., et al. 2013, A\&A 554, 47
\noindent

Grand, R., et al. 2015, MNRAS 447, 4018
\noindent

Hayden, M., et al. 2014, ApJ, 808, 132
\noindent

Jones, T., et al. 2010, MNRAS 404, 1247
\noindent

Jones, T., et al. 2013, ApJ 765, 48
\noindent

Jones, T., et al. 2015, AJ 149, 107
\noindent

Kewley, L. J., \& Ellison, S. L. 2008, ApJ, 681, 1183
\noindent

Liu, C., Xu, Y., Wan, J.-C., et al. 2017, Research in Astronomy and Astrophysics, 17, 096 
\noindent

Magrini, L., et al. 2009, ApJ 696, 729
\noindent

Magrini, L., et al. 2010, A\&A 512, 63
\noindent

Magrini, L., et al. 2016, A\&A 588, 91
\par\noindent

Majewski, S.~R., Schiavon, R.~P., Frinchaboy, P.~M., et al. 2017, AJ, 154, 94
\noindent

Martell, S.~L., Sharma, S., Buder, S., et al. 2017, MNRAS, 465, 3203 
\noindent

Pagel, B. E. J., \& Edmunds, M. G. 1981, ARA\&A 19, 77 
\noindent

Peimbert, M., \& Costero, R. 1969, Boletin de los Observatorios Tonantzintla y Tacubaya, 5, 3
\noindent

Pettini, M., \& Pagel, B. E. J. 2004, MNRAS, 348, L59
\noindent

S\'anchez, S. F., et al. 2012, A\&A, 546, A2
\noindent

S\'anchez, S. F., et al. 2014, A\&A 563, A49
\noindent

S\'anchez-Menguiano, L., et al. 2018, A\&A, 609, A119
\noindent

Searle, L. 1971, ApJ, 168, 327 
\noindent

Stanghellini, L., et al. 2010, A\&A 521, 3
\noindent

Stanghellini, L., \& Haywood, M. 2010, ApJ 714, 1096
\par\noindent

Stanghellini, L., et al. 2014, A\&A 567, 88
\noindent

Stanghellini, L., \& Haywood, M. 2018, ApJ 862, 45
\noindent

Stasinska, G., et al. 2013, A\&A 552, 12
\noindent

Steinmetz, M., Zwitter, T., Siebert, A., et al. 2006, AJ, 132, 1645
\noindent

Wright, S. A., et al. 2014, SPIE 9147, 9
\noindent

Yanny, B., Rockosi, C., Newberg, H.~J., et al. 2009, AJ, 137, 4377 
\noindent

Yuan, T. -T., et al. 2011, ApJ 7232, L14
\noindent

Zhang, K., Zhu, Y., \& Hu, Z. 2016, Proc. SPIE, 9908, 99081P 
\noindent
\end{document}